\documentstyle[12pt]{article}

\newcommand{\resection}[1]{\setcounter{equation}{0}\section{#1}}

\newcommand{\EQ}{\begin{equation}}
\newcommand{\EN}{\end{equation}}

\begin{document}

\topmargin 0pt
\oddsidemargin 5mm

\renewcommand{\thefootnote}{\fnsymbol{footnote}}

\newpage
\setcounter{page}{0}
\begin{titlepage}
\begin{center}
{\large NONPERTURBATIVE EVALUATION OF THE DIFFUSION\\
RATE IN FIELD THEORY AT HIGH TEMPERATURES\\}
\vspace{1cm}
{\large Alexander Bochkarev$^{\, a}$}\footnote{On leave of absence from INR,
Russian Academy of Science, Moscow 117312, Russia}\footnote{Work supported by
Tomalla-Stiftung} {\large and Philippe de Forcrand$^{\, b}$} \\
\vspace{.5cm}
{\em $^{\, a}$Institute for Theoretical Physics, University of Bern \\
Sidlerstrasse 5, CH-3012 Bern, Switzerland.} \\
\vspace{.5cm}
{\em $^{\, b}$IPS, ETH Zurich, CH-8092 Zurich, Switzerland} \\

\end{center}
\vspace{0.2cm}

\begin{abstract}

Kramer's approach to the rate of the thermally activated escape from a
 metastable state is extended to field theory. Diffusion rate in the
 1+1-dimensional
Sine-Gordon model as a function of temperature and friction coefficient
is evaluated numerically by solving the Langevin equation in real time.
A clear crossover from the semiclassical to the high temperature domain
is observed. The temperature behaviour of the diffusion rate allows one
to determine the kink mass which is found equal to the corresponding
classical value. The Kramer's predictions for the dependence on viscosity are
qualitatively valid in this multidimensional case. In the limit of vanishing
friction the diffusion rate is shown to coincide with the one obtained from
the direct measurements
of the conventional classical real-time Green function at finite temperature.

\end{abstract}
\end{titlepage}

\renewcommand{\thefootnote}{\arabic{footnote})}
\setcounter{footnote}{0}

\newpage
\resection{Introduction}
Statistical systems with finite energy barriers separating different
domains of the phase space may exhibit metastable states. Evaluation of the
rate of escape from those states is an important dynamical problem. The
simple example of a
 statistical system where one degree of freedom is exposed to the potential
illustrated by Fig.1 was first considered by Kramers \cite{kram}.
 At temperatures :
\EQ
\omega_{0} \, \ll \,  T \,  \ll \, E_{S}    \label{Tdom}
\EN
where $\omega_{0}$ is the scale of quantum fluctuations and  $E_{S}$ is the
barrier height, the classical thermodynamical fluctuations contribute to the
escape rate $\Gamma \sim \exp (-E_{S}/T)$ of the system initially localized in
the well. Kramers \cite{kram} suggested to determine the escape rate by means
of the classical Langevin equation:
\EQ
\dot{\phi} \, = \, p \, / \, M \, , \;  \; \; \; \dot{p} \, =
\, - \frac{\partial U}
{\partial \phi} \, - \, \gamma p \, + \, \eta (t)   \label{kr}
\EN
\EQ
\left< \eta (t)\, \eta (0) \right> \; = \; 2 T \gamma M \delta (t)
 \label{norm}
\EN
where $\gamma$ is a friction coefficient, introduced as an input parameter in
this phenomenological description and $M$ corresponds to the mass of the
would be Brownian particle. $\gamma$ determines how strong is the
coupling of the system to the heat-bath, represented by the white noise
$\eta$ , and therefore how fast the system reaches equilibrium. The
normalization of the random force (\ref{norm}) (the Einstein relation) comes
from the fluctuation-dissipation theorem.
In the case of one degree of freedom the escape rate $\Gamma (T,\gamma )$
has been obtained analytically \cite{hanggi}
in the domain of very small $\gamma$ and for moderate-to-large friction
$\gamma \gg \omega_{-}$ , where $\omega_{-}$ is the negative eigenmode of
the fluctuations near the stationary point $\phi \, = \, \phi_{s}$.
In quantum field theory one often deals with an analogous problem of
penetration through some energy barrier. One very important example is
anomalous fermion (or axial charge) number violation in gauge theories with
nontrivial structure of the ground state \cite{thooft}. Here the static
energy barrier separates different classical vacua with definite integer
 Chern-Simons numbers \cite{barr}.
Penetration through this energy barrier leads to the dissipation of the
fermionic number. The
system at high temperature is expected to exhibit thermal activation
behaviour of the rate of anomalous fermion number non-conservation \cite{krs}.
The problem specific for the field theory is that the energy barrier
- analogue of $U$ in Eq.(\ref{kr}) - is not known explicitly. This energy
barrier is in the multidimensional configurational space. Only the static
field configuration, corresponding to the top of that barrier is known from
topological considerations \cite{manton} or explicitly in some models
\cite{BS}.

In previous works \cite{bdf1} , \cite{bdf2} we suggested to follow Kramers
in studying the energy barrier in field theory. In \cite{bdf2} we introduced
microscopical classical Langevin equation in real-time in the $1+1$
dimensional
Abelian-Higgs theory and observed the Brownian motion of the Chern-Simons
variable between topologically distinct classical vacua. Our measurements of
the diffusion rate revealed its thermal activation dependence on the
temperature in the domain (\ref{Tdom}). Although this approach is applicable
in the interesting domain of high temperatures, where the semiclassical
approximation is not valid, it was difficult to get data there, because it
required rather big lattices. Meanwhile, it was argued in \cite{SB}
that our results were difficult to interpret, since they were based on the
first-order Langevin equation, derived normally for large friction
coefficients, while, in fact, $\gamma \, = \, 1$  had been used.

In the present paper we deal with a simpler model, the Sine-Gordon field
theory in $1+1$  dimensions, which allows one to obtain not only accurate
quantitative results but also to check qualitative conclusions valid beyond
this particular model. In fact, this model is very similar to the Abelian Higgs
theory. On the classical level the
Sine-Gordon theory has an infinite
number of degenerate vacua with finite energy barrier separating them. At
nonzero temperatures one observes a random walk between those vacua. At
temperatures (\ref{Tdom}) this random process goes through the formation of
kink-antikink pairs, so the rate of the process (diffusion rate) is
sensitive to the density of kinks. Previous studies of the density of
kink gas as a function of temperature in the $1+1$ dimensional scalar
field theory with the double well potential revealed an intriguing fact :
the thermal activation behaviour of the density of kinks
${\em n} \sim \exp (-E_{K}^{eff}/T)$ involved an effective kink mass
$E_{K}^{eff}$
which was found to be $20 \div 30 \%$ smaller than the classical value
$E_{K}^{cl}$ \cite{bdf1}, \cite{SB}. One source of uncertainties in
measuring the density of kinks was the ${\em ad hoc}$ criterion
used in counting the
number of kinks in a given field configuration. This criterion was suggested
in \cite{grig}. One advantage of the Sine-Gordon model is the possibility
to avoid this criterion problem. The quantity we measure is very well
defined: it is the diffusion parameter of the average field.
We measure the diffusion rate in the semiclassical domain (\ref{Tdom})
and beyond. The measurements allow us to extract the value of the kink mass
$E_{K}^{eff}$, which we find to coincide with the classical kink mass
$E_{K}^{cl}$. We therefore conclude that the previously observed discrepancy
was an artifact of the criterion used.

 We also measure the viscosity dependence of the rate by means of the
second order Langevin equation. We find a remarkable coincidence of this
dependence with the one predicted by the first order Langevin equation down
 to very small values of the friction coefficient. This implies that one can
use the first order Langevin approach in field theory, in particular, for
$\gamma \, \sim  \, O(1)$.

In the limit of vanishing friction we find a nonzero diffusion rate. Since in
this limit the Langevin equations (\ref{kr}) become the Hamiltonian ones,
we perform a direct numerical measurement of the classical real-time two-point
 Green function, which describes the diffusion, as a Gibbs's average. These
ensemble averages give the same thermal activation behaviour of the rate as
obtained in the Langevin measurements and, what is less trivial, they
prove to be equal in magnitude to the Langevin measurements
 in the limit of vanishing friction. This establishes an important equivalence
between the Gibbs ensemble measurements of the real time classical Green
functions at finite temperature and the corresponding Langevin measurements
done in the asymptotic domain of very small friction.

In the next section we explain the relevance of simulations of the
classical field theories to the behaviour of quantum field theories
at high temperatures. We emphasize a particular relation between the
lattice spacing of the classical systems under consideration and the
temperature. In section 3 we briefly introduce the $1+1$ dimensional
Sine-Gordon model at finite temperature. Then we describe the first- and
second-order Langevin equations. Sections 4 and 5 contain the results of our
numerical simulations.

\resection{Classical systems - for quantum theories}

It is well known that the partition function of a $D+1$-dimensional
quantum field theory in the limit of high temperature may be obtained
by means of the corresponding $D$-dimensional classical field theory.
To see how this emerges consider, for instance, a scalar field theory
in $D+1$ - dimensions, defined by the action of the general form:
\EQ
S = \int d^{D+1}x\left(\frac{1}{2}(\partial\phi)^{2}+
\frac{m^2}{2}\phi^{2} + \lambda U_{int}\right)     \label{Sin}
\EN
The partition function of the corresponding statistical system is
given by functional integral with the following Matsubara action:
\EQ
S_{M} = \int_{0}^{\beta} dx_{4} \int d^{D}x \left(
\frac{1}{2}(\frac{1}{\hbar} \partial_{4} \phi)^{2} +
\frac{1}{2}(\nabla \phi)^{2} +
\frac{m^2}{2}\phi^{2} + \lambda U_{int}\right)
\label{m}
\EN
where $\beta = 1/T$ is inverse temperature, $\hbar$ is Planck
constant and periodic boundary
conditions are imposed on the field in the imaginary time direction.
Eq.(\ref{m}) implies that in any of the limits :
\EQ
\hbar \rightarrow 0  \; , \; \; or \; \; T  \rightarrow \infty  \label{L}
\EN
the static $x_{4}$-independent field configurations dominate in the
functional integral. If one ignores for a moment ultraviolet divergences
and performs the limit (\ref{L}) formally, the contribution of the static
modes is determined by the action
\EQ
S_{eff} = \beta \int d^{D}x \left( \frac{1}{2}(\nabla \phi)^{2} +
\frac{m^2}{2}\phi^{2} + \lambda U_{int}\right)
\label{Seff}
\EN
This action does not mention Planck constant. The functional integral with the
 action (\ref{Seff})
determines the partition function of the corresponding classical system.
To be more accurate, one should keep in mind that nonstatic
modes do not decouple in the ultraviolet divergent diagrams, which means
that the action (\ref{Seff}) involves renormalized Largangian parameters -
running coupling constants, normalized on the temperature.
The renormalization effects are the memory about the quantum nature of the
original theory (\ref{m}). To obtain the effective action (\ref{Seff}) is,
in general, a separate problem \cite{Nadk}. In some simple cases in low
dimensions the action (\ref{Seff}) is simply a static Hamiltonian.

Let us see explicitly how the reduction takes place, for example, in the case
of weak interaction ($\lambda \ll 1$). The free energy of a gas of
particles of mass $m$, corresponding to (\ref{Sin}), in the one-loop
approximation reads as:
\EQ
F = TV\sum_{n} \int \frac{dk^{D}}{(2\pi)^{D}} \, \frac{1}{2}
\, \log \left(\omega_{n}^{2} + \omega_{k}^{2}\right)
\EN
where $\omega_{n}=2\pi n T, n = 0,\pm1,\pm2,...$ are Matsubara frequencies,
$\omega_{k} = \sqrt{m^2 + k^2}$ and $V$ is $D$-dimensional volume.
The temperature dependent part
\EQ
F = TV \int \frac{dk^{D}}{(2\pi)^{D}}
\log \left(1 - \exp{(-\beta \omega_{k})} \right)         \label{FT}
\EN
becomes in the limit $T \gg m$ :
\EQ
F = - TV \int \frac{dk^{D}}{(2\pi)^{D}} \; \beta k n_{B} (\beta k)
\label{Ffin}
\EN
where $n_{B}$ is Bose distribution so that
\EQ
F = - \kappa \, V \, T^{D+1} \label{Fnum}
\EN
with
\EQ
\kappa = \frac{\Omega}{(2 \pi)^{D}} \Gamma (D+1) \zeta (D+1)
\EN
where $\Omega = 2 \pi^{D/2} / \Gamma (D/2)$ is a surface of the $D$ -
dimensional
sphere \cite{LL}.

If we want to obtain the same expression by means of the effective theory
(\ref{Seff}) we have to perform the limit $\beta \rightarrow 0$
in the integrand of Eq.(\ref{Ffin}), which gives divergent factor:
\EQ
F = - TV \int \frac{dk^{D}}{(2\pi)^{D}}                    \label{FR}
\EN
This divergent integral counts the number of degrees of freedom of the
classical field theory (the Rayleigh-Jeans divergency). Thus the partition
function of the classical field theory is something which needs to be
defined. Regularization is necessary for this end.

The vacuum energy of the effective theory (\ref{Seff}).
\EQ
F = - TV \int \frac{dk^{D}}{(2\pi)^{D}} \, \frac{1}{2} \, \log G_{\Lambda}
\label{VE}
\EN
is determined by the regularised propagator
\EQ
G_{\Lambda} = 1 / \left( \omega_{k}^{2} +
\omega_{k}^{4} / \Lambda^{2} \right)                     \label{Prop}
\EN
For the temperature dependent cut-off :
\EQ
\Lambda = c T \; , \; \; c \propto O(1)                   \label{cut}
\EN
one recovers the result Eq.(\ref{Fnum}) with $c$ determined by $\kappa$.

This simple exercise illustrates one general and important statement.
The effective $D$ - dimensional classical field theory (\ref{Seff}),
which serves for the calculation of the high temperature limit of the
corresponding $D+1$ - dimensional quantum field theory, has a physical
cut-off, given by Eq.(\ref{cut}).
If we assume lattice regularization for the original theory (\ref{m}),
we conclude that the continuum limit is performed simultaneously with
the limit (\ref{L}). One should perform thermodynamical limit in the functional
integral with the effective action(\ref{Seff}), but not the continuum one.
It is essential that the lattice spacing $a^{-1} \propto \Lambda$ here
is the physical cut-off, unambiguously fixed by the temperature:
\EQ
a^{-1} \propto T \, / \hbar                 \label{a}
\EN
In this sense the original quantum field theory may be studied at high
temperatures by simulating the classical functional integral with the
regularized action (\ref{Seff}) and a temperature dependent cutoff (\ref{a})
\cite{grig}. As soon as nontrivial lattice spacing dependence is observed
in a study of the classical systems at high temperatures the
relation (\ref{a}) is to be taken into account.

\resection{Kramers approach in field theory}
We consider the Sine-Gordon model in $1+1$ dimensions defined by the action:
\EQ
S = \int d^{2}x \left(\frac{1}{2}(\partial \tilde{\phi})^{2} \, - \,
\frac{m^2}{\lambda}  \, (1-cos[\sqrt{\lambda} \tilde{\phi} (x)])\right)
\label{SS-G}
\EN
On the classical level the theory has an infinite number of degenerate vacua
with $\phi=2\pi n, n = 0,\pm1,\pm2,..$ . In the quantum theory the
mass $m$ determines the scale of quantum fluctuations
and the self-coupling constant $\lambda$ is bounded \cite{col}:
$\lambda < 8 \pi$. The well-known static solution to the classical equations
of motion interpolating between different classical vacua
$\tilde{\phi_k} = \frac{4}{\sqrt{\lambda}}\arctan\left(e^{mx}\right)$
has energy
\EQ
E_k = 8m / \lambda                \label{Mkink}
\EN
which is parametrically large, in the sense that $\lambda$ is
a parameter of the semiclassical (loop) expansion. If $m$ is chosen as a
unit of measure, the Lagrangian in Eq.(\ref{SS-G}) may be rescaled to the one
which has no free parameters:
\EQ
S = \frac{1}{\lambda}  \; \int d^{2}x\left(\frac{1}{2}(\partial \phi)^{2}
\, - \, (1-cos[\phi(x)]) \right)            \label{Srescaled}
\EN

The Sine-Gordon model provides one with a good opportunity to study thermal
activation phenomena in field theory. The potential energy is completely
bounded at infinity for the zero mode $\bar{\varphi} \, = \, \frac{1}{L}
\int_{0}^{L} dx \varphi (x)$, which  allows for the Brownian motion of
 $\bar{\varphi}$ at finite temperature.
Any initial state localized around one classical vacuum is far from
equilibrium.
In the case of periodic boundary conditions
fluctuations of the field between different vacua proceed via formation
of a pair of kink and antikink and their separation from each other.
Creation and annihilation of kink-antikink pairs is essentially a random walk
between different classical vacua. Thus the diffusion rate of
$\bar{\varphi}$ is sensitive to the kink mass, which determines the height of
the static energy barrier between different vacua. For periodic boundary
conditions kinks and antikinks appear in pairs, so
one expects in the semiclassical domain of temperatures
$m \ll T \ll E_{k}$ :
\EQ
\Gamma_{PBC} \; \sim \; exp(\, - \,2 E_k/T)          \label{rateP}
\EN

In the case of free boundary conditions a configuration with a single kink
 may appear as
a result of the thermal fluctuations. The kink's motion through space brings
 the
system from the neighborhood of one classical vacuum to the vicinity of
another vacuum. Therefore for free boundary conditions we expect:
\EQ
\Gamma_{FBC} \; \sim \; exp(\, - \, E_k/T)            \label{rateF}
\EN

Following Kramers, we describe evolution of some initial non-equilibrium
state localized around one classical vacuum by solving the real time Langevin
equations.

Let us discretize the system of size $L \, = \, Na$ , where $a$ is lattice
spacing. The Hamiltonian, corresponding to Eq.(\ref{Srescaled}) is :
\EQ
\tilde{H}\; =\; \tilde{K}\, +\, \tilde{U} \label{ham}
\EN
\EQ
\tilde{K} \; = \; \sum_{n=1}^{N} \; \frac{1}{2} \, \frac{\lambda}{a^{D}}
\,  (\tilde{p}_{n})^{2}
\nonumber
\EN
\EQ
\tilde{U}\; =\; \sum_{n=1}^{N} \;\frac{a^{D}}{\lambda} \,\left\{ \,\frac{1}{2}
 \left( \phi_{n+1} - \phi_{n} \right)^{2} / a^{2} \, +
\, 1-cos[\phi_{n}] \, \right\}
 \label{Hl}
\EN
where $D$ is dimensionality of space, $D=1$ in our case.
To generate a system with the density matrix $\hat{\rho}\; =\;
\exp \left( - \tilde{H}/T \right)$ , one naturally employs the second order
Langevin equations of the form:
\EQ
\dot{\phi}_{n} \; =\; \frac{\lambda}{a^{D}} \, \tilde{p}_{n} (i)
       \label{veloc}
\EN
\EQ
\dot{\tilde{p}}_{n} \; = \; \frac{a^{D}}{\lambda} \, \left\{ \, \left(\phi
_{n-1} \,+ \, \phi_{n+1} \,-\, 2 \phi_{n} \right) /
 a^{2} \,  - \, \sin \left( \phi_{n}\right) \right\}
\,-\, \gamma \tilde{p}_{n} \, + \, \acute{\eta}_{n}   \label{LE}
\EN
which is a straightforward generalization of Eqs.(\ref{kr}), (\ref{norm}) to
the case of many degrees of freedom. Eqs.(\ref{ham})-(\ref{LE}) indicate that
formally we deal with a classical gas of particles of mass:
\EQ
M\;=\; a^{D}/\lambda       \label{mass}
\EN
from which one concludes that the normalization of the random force,
fixed by the Einstein relation (see Eq.(\ref{norm})), is :
 $\left< \acute{\eta}_{n} (t)\, \acute{\eta}_{m} (0) \right> \; =
\; 2 T \gamma a^{D} \delta (t) \delta_{nm} / \lambda$ .
The average kinetic energy is independent of the effective mass $M$
according to the usual formulae of quantum statistics $<\tilde{K}>\,=\,NT/2$.

The form of the rescaled action Eq.(\ref{Srescaled}) implies that the classical
dynamics does not depend on $\lambda$. To see it explicitly rescale the
momentum: $\tilde{p}\,=\,\lambda p$.  Then the density matrix looks like :
\EQ
\hat{\rho}\; =\; \exp \left( - H_{eff}/\theta_{eff} \right) \label{clsysb}
\EN
\EQ
\theta_{eff}\; =\; \frac{\lambda \, T}{a^{D}}       \label{theta}
\EN
\EQ
 H_{eff}\; =\; \sum_{n=1}^{N} \, \left\{ \, \frac{1}{2}(p_{n})^{2} \, + \,
\frac{1}{2} \left( \phi_{n+1} - \phi_{n}
\right)^{2} / a^{2} \, + \, \, 1-cos[\phi_{n}] \, \right\} \label{clsyse}
\EN
The corresponding Langevin equations indeed do not contain the parameter of
the semiclassical expansion $\lambda$ explicitly:
\EQ
\left( \phi_{n} (i+1) \, - \, \phi_{n} (i) \right) /
\varepsilon \; =\; p_{n} (i)  \label{langeq1}
\EN
\begin{eqnarray}
\left( p_{n} (i+1 ) \,-\, p_{n} (i)
\right) /\varepsilon \; &=& \left( \phi_{n-1} \,+ \, \phi_{n+1}
\,-\, 2 \phi_{n} \right) / a^{2} \nonumber \\
&& -\,\sin (\phi_{n}) \,-\, \gamma p_{n} \, + \, \eta_{n} (i)   \label{Leq}
\end{eqnarray}
where :
\EQ
\left< \eta_{n} (i)\, \eta_{m} (j) \right> \; =
\; \frac{2 \gamma \theta_{eff} }{\varepsilon} \delta_{nm} \delta_{ij}
   \label{noise}
\EN
and we have discretized Langevin time by the amount of $\varepsilon$.
Physical quantities are obtained in the limit $\varepsilon \rightarrow 0$.
Notice that the normalization of the white noise (\ref{noise}) is different
from that in \cite{SB}.

Eqns.(\ref{clsysb})-(\ref{clsyse}) define the effective classical system we
are going to
simulate. It incorporates the lattice spacing $a$ as an input parameter,
determined in the underlying quantum theory by Eq.(\ref{a}).
The kinetic energy of the effective system (\ref{clsysb})-(\ref{clsyse})
averaged over the evolution (\ref{Leq}) may be evaluated to check  the
effective temperature $\theta_{eff}$:
\EQ
\langle K_{eff} \rangle \; =\; \frac{1}{2} \, N \, \theta_{eff}
   \label{Kin}
\EN

One may, of course, consider the first-order Langevin equation instead
of the second-order one. In the Lagrangian formalism, both sets of equations
(\ref{ham})-(\ref{LE}) and
(\ref{clsysb})-(\ref{clsyse}) are equivalent to the following second-order
Langevin equation for the field $\phi$:
\EQ
\ddot{\phi}_{n}\, + \,\gamma \dot{\phi}_{n} \; = \; -
\frac{\partial H_{eff}}{\partial \phi_{n}} \, + \,\eta_{n}
\label{laglang}
\EN
The second term in the l.h.s. of Eqn. (\ref{laglang}) is the damping force.
It dominates over the first one at large friction or large times $t \, > \,
\gamma^{-1}$ . Then neglecting the first term in the l.h.s. of
Eq.(\ref{laglang}) one obtains the first order Langevin equation:
\EQ
\phi_{n} (i+1) \, - \, \phi_{n} (i) \; = \;- \, \frac{\varepsilon}{\gamma} \,
\frac{\partial H_{eff}}{\partial \phi_{n}} \; + \;
\sqrt{\frac{2 \, \theta \,  \varepsilon}{\gamma \, a^{D}}} \; \xi_{n}
   \label{firlan}
\EN
where $\theta  = \lambda T$ and $\xi$ is a random variable with
Gauss's distribution of variance $1$. $\theta$ is the only parameter of the
classical model (apart from the lattice spacing). To fix the physical
temperature one needs to know $\lambda$. Since $\lambda$ is a parameter of the
semiclassical expansion according to Eqn.(\ref{Srescaled}), it is fixed
only in the underlying quantum theory.

The r.h.s of Eqn.(\ref{firlan})
depends on the ratio $\varepsilon / \gamma$, not separately on $\varepsilon$
and $\gamma$. This implies a simple friction dependence of, say, the diffusion
rate determined by means of the first-order Langevin equation: the diffusion
rate must be inversely proportional to the friction coefficient
$\Gamma (T,\gamma ) \, \sim \, 1/\gamma $. Thus
within the first-order Langevin treatment the absolute value of the diffusion
rate is fixed in a simple way by the friction coefficient $\gamma$, which is a
 dimensional quantity. The second-order Langevin equation, valid for
arbitrary friction, predicts nontrivial friction dependence, which has been
evaluated explicitly by Kramers \cite{kram} in the case of one degree of
freedom coupled to the heat bath. In the limit of large $\gamma$ that
friction dependence of the escape rate, of course, reduces to the one we
just derived from the first-order Langevin equation. The question we would
like to clarify is how large the friction coefficient must be in the case of
field theory to allow one to use the first-order Langevin equation instead of
the second-order one. Our numerical data proves to be helpful for this end.

\resection{Numerical simulations}

We have solved the second-order Langevin equations (\ref{langeq1}-\ref{noise})
numerically for the system of size $L \, = \, 50, a \, = \, 1$. This volume is
sufficient to accomodate a few kinks  (which size is $1$ in our units) at low
temperatures. The diffusion parameter $\Delta(t)$ has been measured,
defined following \cite{bdf2} as:
\EQ
\left< \Delta(t) \right> \; = \; \frac{1}{t_o} \, \int_{0}^{t_o} dt' \;
 \left\{ \,
\bar{\varphi}(t'+t) \, - \, \bar{\varphi}(t') \, \right\}^2
 \label{delta}
\EN
The system was observed over $5\cdot10^{7}$ to $2\cdot10^{8}$ iterations
(i.e. a time $t_{o} \, \sim \, 10^6$).  The longer runs were necessary to
reduce statistical errors at the lower temperatures where the diffusion
rate is smaller.
The time step was typically $\varepsilon \, = \, 0.02$ . The Brownian motion
of $\bar{\varphi}$ has been unambiguously observed. Fig.2 demonstrates the
diffusion
at temperature $T \, = \, 6$ for 3 different values of the lattice spacing. The
rate is obviously insensitive to the lattice spacing and the relatively large
value
$a \, = \, 1$ may be used to obtain physical results. Large values of
$a$ are preferable to save computer time.

Periodic as well as free boundary conditions were studied.
In both cases we measured the diffusion rate in
the range of temperatures $\theta \, \cong \, 1.33 \, \div \, 18$. The
semiclassical domain corresponds to $\theta \, \ll \, 8$ in accordance with
(\ref{Mkink}). Fig.3 shows the temperature dependence of the diffusion rate
on a logarithmic scale. In the case of free boundary conditions all the data
points starting from $T \, \cong \, 3$ downwards lie on a straight line. This
implies the expected thermal activation behaviour (\ref{rateF}). The slope of
the straight line is seen to be $8$, which is exactly the classical value of
the kink mass (\ref{Mkink}). In the case of periodic boundary conditions, a
somewhat more interesting phenomenon is observed: starting around
$T \, = \, 2$ the slope of the straight line changes from $8$ to $16$ , which
corresponds to the classical energy of a kink-antikink pair. This is again in
accordance with the expectations (\ref{rateP}) for periodic boundary
conditions. The sensitivity to the boundary conditions appears in the
lowest temperature domain. Fig.4 shows that at higher temperatures there is no
dependence on the boundary conditions. The cross-over temperature below
which the kink-antikink pair energy is observed in (\ref{rateP}) depends on the
system size $L$. The larger the volume $L$ the smaller that crossover
temperature.

The diffusion rate can also be measured in the  high temperature domain
$\theta \, \geq \, 8$ , where the semiclassical approximation is not valid.
Our numerical results are shown on Fig.5 for both types of boundary conditions.
The diffusion rate does not depend on the choice of boundary conditions as
it is not associated with the production of kink-antikink pairs any more.

At high temperatures the Langevin equation (\ref{firlan}) can be integrated
explicitly, because the noise term, whose contribution is proportional to
$\sqrt{T}$, dominates over the regular force. Then the analytical predition
for the rate is :
\EQ
\Gamma \; = \; \frac{2 \, T}{\gamma \, L}          \label{highTrate}
\EN
The numerical measurements are in perfect agreement with Eqn.(\ref{highTrate}).

The crossover temperature from the semiclassical to the high temperature
domain is seen to be around $T \, = \, 3$. Perhaps surprisingly it is less
than the kink energy (\ref{Mkink}). To determine it more accurately we have
measured the probability distribution of the averaged field $\bar{\varphi}$,
whose logarithm determines the effective potential $V(\bar{\varphi} , T)$.
One can see on Fig.6 that at $T \, = \, 2$ there is a deep parabolic well at
the origin, so that the field most of the time oscillates around
$\bar{\varphi} \, =  \, 0$. Beyond the parabolic well the potential is flat.
This implies the free motion between the different vacuum sectors,
which proceeds via the formation of kink-antikink pairs, seen as the
modulation of the high frequency oscillations corresponding to the parabolic
well. The deeper the well the better is the very notion of the kink
configuration at $T \, \neq \, 0$. At temperature $T \, = \, 3$ the parabolic
well is rather shallow, so the low frequency modulations usually interpreted as
kinks become hard to identify. The effective potential is almost entirely flat,
which means that the transitions between different classical vacua do not
require smooth low frequency configurations any more. This is obviously the
boundary of the semiclassical domain of temperatures. At temperature $T \, = \,
 6$ there is not any sign of the confining parabolic well.

Thus we have found
that the effective potential $V(\bar{\varphi} , T)$ of the diffusing variable
is flat in the high temperature domain, starting from $T \, \simeq \, 3$. This
result is interesting in the context of the baryogenesis within the standard
electroweak theory \cite{shap1}. In the Standard Model there is a variable,
which is expected to diffuse at finite temperature. It is the Chern-Simons
variable $N_{CS}$. Diffusion of $N_{CS}$ leads to the dissipation of the
baryonic number. This effect could be responsible for the production of the
baryon asymmetry in the expanding Universe. In \cite{shap} a scenario for
baryogenesis within the Standard Model was suggested, in which the crucial
assumption was the flat effective potential $V(N_{CS})$ in the high temperature
domain. While to measure properly $V(N_{CS})$ in the Electroweak Theory is not
easy, our data in the Sine-Gordon model indicates that the flat effective
potential $V(N_{CS})$ is a rather plausible assumption. We would like to
emphasize again the similarity between the Sine-Gordon model and Abelian Higgs
theory in 1+1 - dimensions. The zero mode $\bar{\varphi}$ of the Sine-Gordon
theory plays the role of the Chern-Simons variable in the Abelian Higgs model.
One can decompose $\varphi \; \rightarrow \; \bar{\varphi} \, + \, \delta
\varphi$,
to demonstrate that the effective potential $V(\bar{\varphi} , T)$ is a
bounded periodic function, allowing for the Brownian motion.

We now turn to the dependence on the friction coefficient. The friction
coefficient $\gamma$ is a dimensional input parameter in the Langevin equation
(\ref{laglang}). It determines the scale as it is seen, for instance, from
formula (\ref{highTrate}), derived for large $\gamma$. Although the friction
dependence of the escape rate has been obtained analytically in \cite{kram} in
the case of one degree of freedom exposed to the heat bath, it is a less
trivial task in the field theory. We have measured the friction dependence of
the diffusion rate numerically by solving equations (\ref{langeq1}) -
(\ref{Leq}) for the two temperatures $T \, = \, { 2, 6 }$ and for various
friction coefficients. The results are shown in Fig. 7, 8. In both the
semiclassical
 and high temperature domains we find the same behaviour of the rate. The
inverse rate varies linearly with the friction coefficient from large
$\gamma \, > \, 1$ down to very small $\gamma \, \sim \, 10^{-2}$. This simple
friction dependence coincides with the one predicted by the first order
Langevin equation. Therefore we obtain experimental evidence that the first
order Langevin equation may be used to explore the large time behaviour of the
correlators not only for large, but also for small friction coefficients, in
particular, for $\gamma \, \sim \, 1$.

One can see from Fig. 8 that the rate does not diverge in the limit $\gamma
\, \rightarrow \, 0$. It approaches some finite asymptotic value. Meanwhile
the original Langevin equations become the conventional Hamiltonian ones in the
 limit $\gamma \, \rightarrow \, 0$. Therefore we expect the rate
extracted from the second order Langevin simulations in the limit of vanishing
friction to coincide with the one obtained by means of microcanonical
simulations. To verify this conjecture we first of all notice that the
microcanonical simulations of classical systems in real time \cite{grig}
naturally correspond to some fixed energy, not temperature. To obtain the
time dependent Gibbs averages, one has to average the result of the
microcanonical measurement over the initial field configuration with the
Gibbs weight:
\EQ
\left< \Delta(t) \right>\;=\;\frac{\int {\cal D} \varphi_{0} {\cal D} p_{0} \;
exp\left( \, -\, H(\varphi_{0},p_{0}) \, / \, T \, \right) \;
\left\{\, \bar{\varphi}(\, t,\, \varphi_{0} \, , \, p_{0}\, ) \, -
\, \bar{\varphi}(0) \, \right\}^{2} }
{\int {\cal D} \varphi_{0} {\cal D} p_{0}\;exp\left( \,-\,H(\varphi_{0},p_{0})
\, / \, T \, \right) }                  \label{gibbs}
\EN

 The problem here is that the value of the diffusion parameter
$\Delta$ at time $t$, obtained as a result of the Hamiltonian evolution, is a
functional of the initial field configuration
$\{\varphi_{0}, p_{0}\} \, = \, \{\varphi(t=0), p(t=0)\}$. This functional
is not known explicitly. However, following the suggestion by J.Smit \cite{jan}
we evaluate this functional numerically. This means that to measure the
Gibbs averaged diffusion parameter directly we calculate numerically
$(\varphi(t) \, - \, \varphi(0))^{2}$ from the Hamiltonian evolution for
a given starting configuration $\{ \varphi_{0}, p_{0} \}$, then update this
starting configuration by means of the Metropolis procedure, corresponding to
 some temperature $T$, then do microcanonical evolution again and so perform
the Gibbs average.

In this way the diffusion parameter (\ref{gibbs}) has been measured. The
function
$\left< \Delta (t) \right>$ is shown on Fig. 9. One can see a crossover between
  two domains of time. At short times $t \, \leq \, O(10^{2})$ the
deterministic behaviour is observed $\left< \Delta (t) \right> \,\sim\, t^{2}$,
while at $t \, \geq \, O(10^{2})$ the Brownian motion sets in
 $\left< \Delta (t) \right> \,\sim\, t$.
The corresponding diffusion rate as a function of temperature is shown on
Fig. 10. The thermal activation behaviour is clearly seen at low temperatures.
As a result we observe a remarkable coincidence between the data from the
second order Langevin simulations in the limit of vanishing friction and
the direct Gibbs averages. One obtains $\Gamma(\, T=2 \,) \,\simeq \, .3$ and
$\Gamma(\, T=6 \,) \,\simeq \, 8.$ from both Fig. 8 and Fig. 10. This result
answers the question of how to relate the Langevin measurements to the direct
calculations of the Gibbs's average (\ref{gibbs}).

\resection{Comments on the errors}
As mentioned above all the runs which results are presented were long enough
to make statistical errors negligible. The two main sources of systematic
errors are the finite lattice spacing $a$ and the time step $\varepsilon$. The
insensitivity to the lattice spacing was discussed before (see Fig. 2).
 Here we would like to address the artifacts of the discretization of the
Langevin time.

It is known \cite{Teff} that the finite time step in Langevin simulations
makes the temperature of the simulated system $T_{eff}$ larger than the input
one $T$: $\Delta T \, \equiv \,  T_{eff} - T \, \sim \, O(\varepsilon)$. To
check the effective temperature we measure the diffusion parameter at rather
small times, as shown on Fig. 11. The Brownian diffusion is clearly observed
at short times, but it has nothing to do with the motion of the system between
topologically different vacua. At short times the white noise always dominates
over the regular force in the Langevin equation. Therefore the short time
behaviour of $\left< \Delta (t) \right>$ is, in fact, controlled by the free
Langevin equation, which immediately implies diffusion of $\bar{\varphi}$. This
short time diffusion is already sensitive to the discretization effects. Its
diffusion rate is given by $T_{eff}$, which has been measured numerically
and is shown on Fig. 12. One can see that the effective temperature of the
simulated system follows the input one with fairly high accuracy. Since these
measurements are obtained from very short runs, the computer time needed is
 quite small. So we have also checked that $\Delta T \, = \, c \varepsilon$,
with $c \, = \, 1.5 \pm .5$.

Another alternative to check the effective temperature of the simulated system
is to measure the Langevin average of the canonical momentum squared over the
whole run: $<p^{2}> \, = \, T_{eff}$. This way is not easily generalizable for
theories where there is a coupling between the canonical momenta and
coordinates in the Hamiltonian, like in gauge theories.

\section{Conclusions}
The study of the Sine-Gordon field theory in 1+1-dimensions shows that the
classical
Langevin simulations in real time prove to be efficient in obtaining valuable
information about the nonperturbative effects in field theory at high
temperatures. In the semiclassical domain of temperatures kinks are seen
as smooth modulations of the high frequency oscillations of the field.
Diffusion between different classical vacua is due to random process of
production and free motion of kink-antikink pairs during the Langevin
evolution. This is very well confirmed by the
measured temperature dependence of the diffusion rate, which exhibits
thermal activation behaviour with the classical value of the kink mass in the
 Boltzmann exponent.

Crossover is observed between the semiclassical and high-temperature
domain, where the semiclassical approximation is not valid. The effective
potential of the diffusing variable is found flat at high temperatures.
In the high temperature domain at moderate friction the diffusion rate
follows prediction of the free Langevin equation.

The dependence of the diffusion rate on the friction coefficient is found
identical to the  prediction of the first order Langevin equation down to
rather small friction coefficients $\gamma \, \sim \, 10^{-2}$. This
justifies the use of the first order Langevin equation in the calculations
of correlation functions at large real times.

We also measure the diffusion parameter directly as the Gibbs ensemble average
by means of the Metropolis procedure and microcanonical evolution. The
corresponding diffusion rate is found to coincide with the smooth
extrapolation of Langevin
measurements in the limit of vanishing friction coefficient.

\resection{Acknowledgements}
Fruitful discussions with P. van Baal, O. Lanford, H. Leutwyler,
P. Hasenfratz, J. Hetrick, J. Smit are gratefully acknowledged.


\newpage
{\large Figure Captions}\\

Fig. 1 Typical potential creating a metastable state at finite temperature.

Fig. 2 Time dependence of the diffusion parameter (\ref{delta}) obtained from
the second order Langevin equation.
Measurements were performed every 100 iterations. The parameters are:
$T = 6, L = 50, \varepsilon = .02$. Three solid lines correspond to
different lattice spacings: $a \, = \, \{1, .5, .25\}$ ; the dashed line is
a fit with slope 1.

Fig. 3 Temperature dependence of the diffusion rate on a logarithmic scale
for: $a)$ free, $b)$
periodic boundary conditions. $L = 50,\, a = 1$. Typical uncertainty is
shown for $T = 2$. The dashed lines correspond to the Boltzmann exponent
with slope given by the classical value of the kink mass $M_{k} = 8$ in
$a)$ and the energy of the kink-antikink pair $2M_{k} = 16$ in $b)$.

Fig. 4 Temperature dependence for free vs. periodic boundary conditions
at both high and low temperatures.

Fig. 5 Temperature dependence of the diffusion rate at high temperatures.
The dashed line is a prediction from the corresponding first order free
Langevin equation.

Fig. 6 Logarithm of the probability distribution (effective potential)
of $\bar{\varphi}$ at different temperatures indicated on the plots.

Fig. 7 Friction dependence of the rate in the low ($a$) and high ($b$)
temperature domains obtained from the second order Langevin equation.
$L = 50, \, a = 1$. The dashed straight line is a prediction from the first
order Langevin equation, derived normally for large friction $\gamma > 1$ or
large times.

Fig. 8 Same as in Fig. 7, but only the points corresponding to very small
values of the friction coefficient are shown. The cross-points on the y-axis
are the extrapolations.

Fig. 9 Time dependence of the diffusion parameter (\ref{delta}) obtained from
direct measurements of the Gibbs average (\ref{gibbs}) for periodic
boundary conditions. Temperature is $T = 1.5$. The initial field configuration
was updated by means of the Metropolis procedure 100 times and each time
$10^{6}$ leap-frog iterations of the microcanonical evolution were performed.
$L = 50,\, a = 1, \, \varepsilon = .1$

Fig. 10 Temperature dependence of the diffusion rate obtained from
direct measurements of the Gibbs average (\ref{gibbs}).
The dashed line corresponds to the Boltzmann factor with
the energy of the kink-antikink pair. The parameters of the measurements are
as in Fig. 9.

Fig. 11 The typical diffusion at short times, controlled by the free Langevin
equation.

Fig. 12 The effective temperature of the simulated system obtained from the
short time diffusion. The dashed line is a fit. The solid line corresponds to
the input temperature. The deviation is proportional to the time step
$\varepsilon$.

\end{document}